\documentclass{sig-alternate-10pt}
\usepackage{epsfig,endnotes}

\usepackage{tikz}
\usepackage{pgfplots}
\usepackage[most]{tcolorbox}
\usepackage{paralist}
\usepackage{multirow}
\usepackage{todonotes}
\usepackage{url}
\usepackage{graphicx}
\usepackage{adjustbox}
\usepackage{xspace}
\usepackage{verbatim}
\usepackage{amssymb}
\usepackage{pifont}
\usepackage{balance} 
\usepackage{paralist}
\usepackage{subcaption}

\newcommand{\YA}[1]{\textcolor{green}{[YA: #1]}}
\newcommand{\GS}[1]{\textcolor{brown}{[GS: #1]}}   
\newcommand{\KK}[1]{\textcolor{purple}{[KK: #1]}}  

\pgfplotsset{compat=1.14}
\graphicspath{ {images/} }

\begin{document}
%
\title{\Large \bf PrivacyProxy: Leveraging Crowdsourcing and In Situ Traffic Analysis to Detect and Mitigate Information Leakage}


\author{
{Gaurav Srivastava}\\
{Technical University of Munich}\\
{ga79daw@tum.de}
\and 
 {Kunal Bhuwalka}\\
{Carnegie Mellon University}\\
{kbhuwalk@andrew.cmu.edu}
\and
 {Swarup Kumar Sahoo}\\
{Carnegie Mellon University}\\
{swarupks@andrew.cmu.edu}
\and
 {Saksham Chitkara}\\
{Carnegie Mellon University}\\
{schitkar@andrew.cmu.edu}
\and
 {Kevin Ku}\\
{Carnegie Mellon University}\\
{kku@alumni.cmu.edu}
\and
 {Matt Fredrikson}\\
{Carnegie Mellon University}\\
{mfredrik@cs.cmu.edu}
\and
 {Jason Hong}\\
{Carnegie Mellon University}\\
{jasonh@cs.cmu.edu}
\and
 {Yuvraj Agarwal}\\
{Carnegie Mellon University}\\
{yuvraj.agarwal@cs.cmu.edu}
}


\maketitle


%




\def\PP{PrivacyProxy\xspace} 
\def\totalAppsFieldStudy{191\xspace}	 
\def\totalPublicSignaturesFieldStudy{1803\xspace}
\def\totalApps{500\xspace}	 
\def\totalAppsForMoreTraining{40\xspace}
\def\totalPublicSignaturesControlled{12603\xspace}
\def\appswithCertificatePinning{22\xspace}
\def\appswithLogin{20\xspace}
\def\appswithnonetwork{6\xspace}
\def\totalAppsRunOnDroidbot{452\xspace}
\def\PIILeaksControlled{18407\xspace}

\def\SignCount{1803\xspace} 
\def\AppCount{430\xspace} 

\def\PrecisionImprovement{20\%\xpace} 
\def\RequestCanNotParse{0.5988\xpace} 

\def\PrecisionControlled{61.1\%\xspace} 
\def\RecallControlled{100\%\xspace}
\def\FScoreControlled{0.759\xspace}
\def\TPControlled{3184\xspace}
\def\FPControlled{2021\xspace}
\def\FNControlled{0\xspace}
\def\TNControlled{13202\xspace}

\def\PrecisionAdditionalTrainingStart{63.6\%\xspace} 
\def\RecallAdditionalTrainingStart{100\%\xspace}
\def\FScoreAdditionalTrainingStart{0.778\xspace}
\def\TPAdditionalTrainingStart{321\xspace}
\def\FPAdditionalTrainingStart{184\xspace}
\def\FNAdditionalTrainingStart{0\xspace}
\def\TNAdditionalTrainingStart{1803\xspace}

\def\PrecisionAdditionalTraining{76.4\%\xspace} 
\def\RecallAdditionalTraining{100\%\xspace}
\def\FScoreAdditionalTraining{0.866\xspace}
\def\TPAdditionalTraining{321\xspace}
\def\FPAdditionalTraining{99\xspace}
\def\FNAdditionalTraining{0\xspace}
\def\TNAdditionalTraining{1888\xspace}

\def\PrecisionFieldStudy{79.3\%\xspace} 
\def\RecallFieldStudy{100\%\xspace}
\def\FScoreFieldStudy{0.885\xspace}
\def\TPFieldStudy{92\xspace}
\def\FPFieldStudy{24\xspace}
\def\FNFieldStudy{0\xspace}
\def\TNFieldStudy{XX\xspace}

\def\PerformanceOverhead{14.2\%\xspace}
\def\BatteryLoss{XX\%\xspace}
\def\UsersWithNoBatteryLoss{85\%\xspace}
\def\UsersWithNoNwPerformanceLoss{68\%\xspace}

\def\UploadThresholdExperiment{3\xspace}
\def\UploadThresholdDeployment{10\xspace}

\def\privateProbThreshold{0.95\xspace}


\begin{abstract}
Many smartphone apps transmit personally identifiable information (PII), often without the user's knowledge. To address this issue, we present \textit{PrivacyProxy}, a system that monitors outbound network traffic and generates app-specific signatures to represent sensitive data being shared. \PP uses a crowd-based approach to detect likely PII in an adaptive and scalable manner by anonymously combining signatures from different users of the same app. Furthermore, we do not observe users' network traffic and instead rely on hashed signatures. We present the design and implementation of \PP and evaluate it with a lab study, a field deployment, a user survey, and a comparison against prior work. Our field study shows \PP can automatically detect PII with an F1 score of \FScoreFieldStudy. \PP also achieves an F1 score of \FScoreControlled in our controlled experiment for the \totalApps most popular apps. The F1 score also improves to \FScoreAdditionalTraining with additional training data for \totalAppsForMoreTraining apps that initially had the most false positives. We also show performance overhead of using \PP is between 8.6\% to \PerformanceOverhead, slightly more than using a standard unmodified VPN, and most users report no perceptible impact on battery life or the network. 
\end{abstract}


%


\section{Introduction}
\label{sec:Introduction}
\vspace{-2mm}




NIST defines personally identifiable information (PII) as ``any information that can be used to distinguish or trace an individual's identity'' \cite{NIST_PII}. Prior work has shown that many smartphone apps aggressively collect PII \cite{ProtectMyPrivacy, Almuhimedi_Your_Location, TaintDroid, Liu_Identifying_PII, MAdScope, ReCon, PrivacyGuard}. While PII can be used to provide useful services, it is often also used to track users' behaviors without their consent, potentially by multiple third-party trackers and across different apps \cite{razaghpanah_ndss2018_Apps, Chitkara_Ubicomp2017_ContextAwarePrivacy}. A recent study looked at the evolution and updates to popular apps over time and in fact found that PII being accessed by some apps has actually been increasing in newer versions \cite{ren_ndss2018_bugfixes}, with many apps showing no improvements. Being able to detect PII in a reliable and scalable manner would offer many opportunities for improving privacy, e.g. scrubbing PII before it leaves a user's device or allowing a third party to assign privacy grades to apps based on the pervasiveness and sensitivity of the data being collected.

Focusing on Android, it is well known that any app can access the unique, per-device Android ID \cite{AndroidID} without an Android permission. Apps and third-party libraries can also generate their own unique identifiers. Furthermore, identifiers may not always have predictable names or formats, making it hard to have a comprehensive blacklist based on regular expressions.
Adding to the complexity of protecting PII on Android is the pervasive use of third-party libraries \cite{Chitkara_Ubicomp2017_ContextAwarePrivacy,PrivacyGradePaper}. Lin et al. show that a typical Android app uses 1.59 (std = 2.82) third-party libraries \cite{PrivacyGradePaper}. Libraries within the same app can access the same PII, but for different purposes. For example, an app might contain two libraries that use location, one for maps and the other for ads. This means enforcing PII access control on a per-app basis may not suit the user's needs. 
Furthermore, Chitkara et al. found that the top-100 most popular third-party libraries account for over 70\% of all sensitive data accesses across the apps they tested \cite{Chitkara_Ubicomp2017_ContextAwarePrivacy}, using their app on a rooted smartphone. 
This finding suggests that denying PII access for one app is not necessarily effective, since a given third-party library might be embedded in other apps that the user has.

To address these problems, we present \textit{PrivacyProxy}, a privacy-sensitive approach for automatically inferring likely PII using network analysis and crowdsourcing. Like other proxy-based approaches \cite{ReCon,PrivacyGuard,HayStack}, we use a VPN to intercept and analyze smartphone network data, thus avoiding having to modify apps or make changes to the OS. However, instead of routing all of the user's network traffic to a remote server for analysis (like ReCon \cite{ReCon}), \PP looks for key-value pairs in HTTP requests on the smartphone itself and only sends cryptographically hashed signatures to our servers, thereby minimizing potential privacy and security risks. Each signature represents the content of a request to a remote host without revealing actual content, letting us identify potential PII based on the uniqueness of a signature. Additionally, instead of requiring hard-coded rules or regular expressions (like PrivacyGuard \cite{PrivacyGuard} or HayStack \cite{HayStack}), or requiring users to label network traces for training (like ReCon \cite{ReCon}), our approach is adaptive and robust against some forms of data obfuscation such as hashing a PII or encoding it. 
Furthermore, we show that as \PP observes more network requests, we can classify PII more accurately. Lastly, since the PII detection and signature generation happen on device, \PP also provides controls to the user to allow or block PII at a fine granularity, while maintaining a small trusted computing base. In summary, we make the following contributions:



\vspace{-2mm}
\begin{itemize}
     \item We present a privacy-sensitive approach that uses network analysis and crowdsourcing to detect likely PII leaked by apps without requiring any changes to apps or the OS. Our approach uses a novel and adaptive way of detecting PII without relying on a priori knowledge of names or formats of identifiers. Furthermore, our approach does not require labeled network traces or any user intervention.

   \vspace{-2mm}
    \item We present the results of a controlled experiment, a field deployment, a user survey, and a comparison against two pieces of past work. \PP achieved a F1 score of \FScoreControlled in the controlled experiment (\totalApps apps) and \FScoreFieldStudy in the field deployment (\totalAppsFieldStudy apps). For the \totalAppsForMoreTraining apps with the most false positives, we show that as additional network requests are observed, the F1 score improves to \FScoreAdditionalTraining. We also show the trade-off between precision and recall. Furthermore, we show that \PP finds more kinds of PII than past work. Finally, our user survey shows that 85\% and 68\% of users perceived no change in battery life and network performance respectively. 


    
\end{itemize}

\section{Background and Related work}
\label{sec:Related work}


Many types of data can be characterized as PII, e.g. name, street address, IP address, photos, fingerprints, race, and religion \cite{NIST_PII}. We take a narrow view of PII, focusing primarily on unique identifiers, such as user names, Android ID, IMEI, MAC, Advertising ID, phone numbers, identifiers generated by apps or libraries (e.g. Java's UUID class or hashing one's MAC address), and likely identifiers, such as install time of an app. Here, we limit the scope of identifiers to just single values. 




We group prior work into three categories: (1) code analysis techniques, which aim to improve privacy behavior of mobile apps by analyzing the apps' source code or binaries; (2) approaches that modify the underlying mobile OS; 
and (3) network flow analysis techniques to detect and remove PII. 

\textbf{Code Analysis: } Static analysis approaches identify the uses of sensitive information by analyzing an app's source code or binary. FlowDroid \cite{FlowDroid} and DroidSafe \cite{DroidSafe} are information-flow analysis tools that, given a set of sources and sinks, can statically identify possible PII leaks.  
AndroidLeaks \cite{AndroidLeaks} uses taint-aware slicing to find potential PII leaks. PrivacyGrade \cite{PrivacyGradePaper, PrivacyGrade} infers the purpose behind each permission request for Android apps by decompiling apps and searching for sensitive API calls made by third-party libraries. Wang et.al. \cite{WangPurposeTextMining} used text mining on decompiled apps to infer the purpose of permission use for location and contact lists. These approaches do not change the functionality of the app at runtime and do not provide PII filtering, but rather are used to inform users and developers of potential data leaks. These techniques also fail to work in the presence of dynamic code loading \cite{DynamicCodeLoading} and reflection techniques. Additionally, these techniques can only detect access to sensitive data using well-defined permissions or APIs rather than arbitrary identifiers an app might generate, which we show is quite common. 





\textbf{Instrumenting the OS and APIs: } Another approach is to modify the underlying OS and intercept certain APIs that access private user data. For example, ProtectMyPrivacy \cite{ProtectMyPrivacy,Chitkara_Ubicomp2017_ContextAwarePrivacy} and xPrivacy \cite{XPrivacy} intercept OS API calls requesting sensitive information and alert the user. Mockdroid \cite{MockDroid} modifies the Android OS so it can substitute private data with mock data. TaintDroid \cite{TaintDroid} offers a custom Android version that does dynamic taint-tracking, enabling information flow tracking from sources (requests for sensitive information) to sinks. AppTracer\cite{Micinski2017AppTracer} uses binary rewriting to determine if sensitive resource usages happen due to user interactions in applications or not. While these approaches augment native privacy protection features of iOS and Android, their applicability is limited to rooted or jailbroken devices. PrivacyBlocker \cite{PrivacyBlocker} uses static analysis on app binaries and replaces calls for private data with hard coded shadow data. This approach, however, requires target apps to be re-written and reinstalled and cannot be done at runtime. 

\PP offers a complementary approach, and one that works with 
\textit{unmodified devices with the stock OS, and unchanged Apps}. In addition, \PP only relies on VPN functionality, which is likely to be continually supported in the future versions of these OSes due to its enterprise use. In addition, our approach can detect an expanded set of PII, including those generated by apps themselves, and not just ones that are accessible using well known OS APIs.

\textbf{Network Flow Analysis: } The closest related work is network flow analysis, which seeks to detect (and in some cases remove) PII as it leaves a device. 
\label{sec:PrivacyOracle}
Privacy Oracle uses differential black-box fuzz testing to detect leaks \cite{PrivacyOracle}, feeding different sets of inputs to an app and looking for associated changes in network traffic. Similarly, Agrigento \cite{continella_ndss2017_obfuscation} uses black box differential analysis to observe whether the network traffic of an app changes when certain known types of PII (such as location or an ID) are modified or kept the same. However, these approaches are hard to scale up because there are millions of smartphone apps, and one must generate multiple sets of inputs per app. In contrast, \PP's novel signature-based detection algorithm scales much better and the accuracy of inferences improves as the number of users increase. Also, \PP uses valid network requests based on users' regular use of their apps rather than generating inputs or modifying well-known PII, which increases the possibility of finding more PII due to increased coverage.

AntMonitor uses the Android VPN service to send network traces to their server, while allowing users to disable traffic at the app level, and choose whether all data or just headers are sent \cite{le_c2bid2015_antmonitor}. PrivacyGuard \cite{PrivacyGuard} and Haystack \cite{HayStack} use Android's VPN Service API to intercept network traffic for privacy. However, they both rely on regular expressions to find potential PII, limiting the types of PII they can detect. For example, an app might generate its own identifier or use non-standard encodings. Recon \cite{ReCon} uses a VPN to redirect all user traffic to a proxy in the cloud and uses heuristics to identify potential PII, which could lead to privacy concerns itself. 

In contrast, \PP offers a hybrid approach, where data is processed locally on the user's smartphone and only hashed data is sent to \PP servers. \PP identifies likely PII based on the uniqueness of the data. Additionally, the \PP server does not see any actual user data, mitigating privacy and security concerns that users may have about their sensitive data leaving the confines of their device. Finally, \PP is more robust to changes in formats of the data compared to other approaches. In Section \ref{sec:Evaluation}, we evaluate \PP against both a regular expression based approach (Haystack \cite{HayStack}) as well as a machine learning based approach (Recon \cite{ReCon}) showing that \PP can detect many more PIIs.

\section{System Architecture}
\label{sec:sysarch}

Our overarching objective with \PP is to develop a practical system that increases smartphone users' awareness of and provide control over PII that apps send over the network. To achieve this, we had a number of design goals. First, to maximize utility for everyday users, our solution must work on unrooted devices. Second, to detect a broad range of PII, including known trackable IDs (e.g. UUIDs, MAC addresses, etc), and previously unknown identifiers dynamically generated by specific apps or libraries. Third, to provide users with effective notifications and fine-grained controls to prevent the flow of PII over the network by blocking or anonymizing the data. Fourth, to impose minimal performance overhead on the device while using \PP. Fifth, to be usable in practice by requiring only minimal user interaction for privacy decisions. Finally, to be scalable with increasing number of users and apps. 

\textbf{Threat Model:} 
We assume developers are generally honest and do not try to deliberately obfuscate their activities. If an app uses custom app-layer encryption, non-standard encodings, or indirect ways to track users, we will not be able to detect PII leakage. This is a limitation of all prior work in this space. Additionally, while we handle SSL using a man-in-the-middle approach, like other current approaches, we are unable to handle apps that are certificate pinned. We show in our evaluation certificate pinning is not that prevalent in practice.

\vspace{-1mm}
\subsection{Intercepting Network Traffic} 

To observe what data (and potential PII) is being sent over the network, we need to intercept  traffic on \textit{unmodified} smartphones. We also need to separate traffic into different flows of information, taking into account individual apps, hosts where data is being sent to, etc. To achieve this, we build on the internal Virtual Private Networking (VPN) service provided by most modern smartphone OSes. Traditional VPNs intercept all device traffic and forward it over an encrypted tunnel to another server, which can then process and/or forward the data to the eventual destination. VPNs can operate at multiple layers, including transport layer (SSL/TLS) and network layer (e.g. IPSec). 

We build upon PrivacyGuard \cite{PrivacyGuard}, which does the following: (a) registers a \texttt{Fake VPN service} to intercept all IP packets; (b) implements \texttt{TCP-Forwarders} and \texttt{UDP-Forwarders} that implement TCP and UDP protocols; (c) instantiates a \texttt{LocalServer} on the smartphone which pretends to be the destination for all app-server communication and acts as a man-in-the-middle (MITM) proxy; (d) connects the LocalServer with the actual remote server for each request and does the necessary TCP handshakes; (e) passes responses from the real server to the TCP forwarders via the LocalServer, which in turn delivers the IP packets back to the app. 

To decrypt SSL/TLS, we perform MITM SSL injection. With the user's explicit consent, we add a trusted root certificate on the device that lets \PP sign any certificate for end hosts that individual apps contact, so that the LocalServer process described above can finish the SSL handshake and decrypt packets sent by an app on the device itself before sending the packets through a new SSL connection with the actual server.


\begin{figure}
\centering
\includegraphics[width=\columnwidth-10pt]{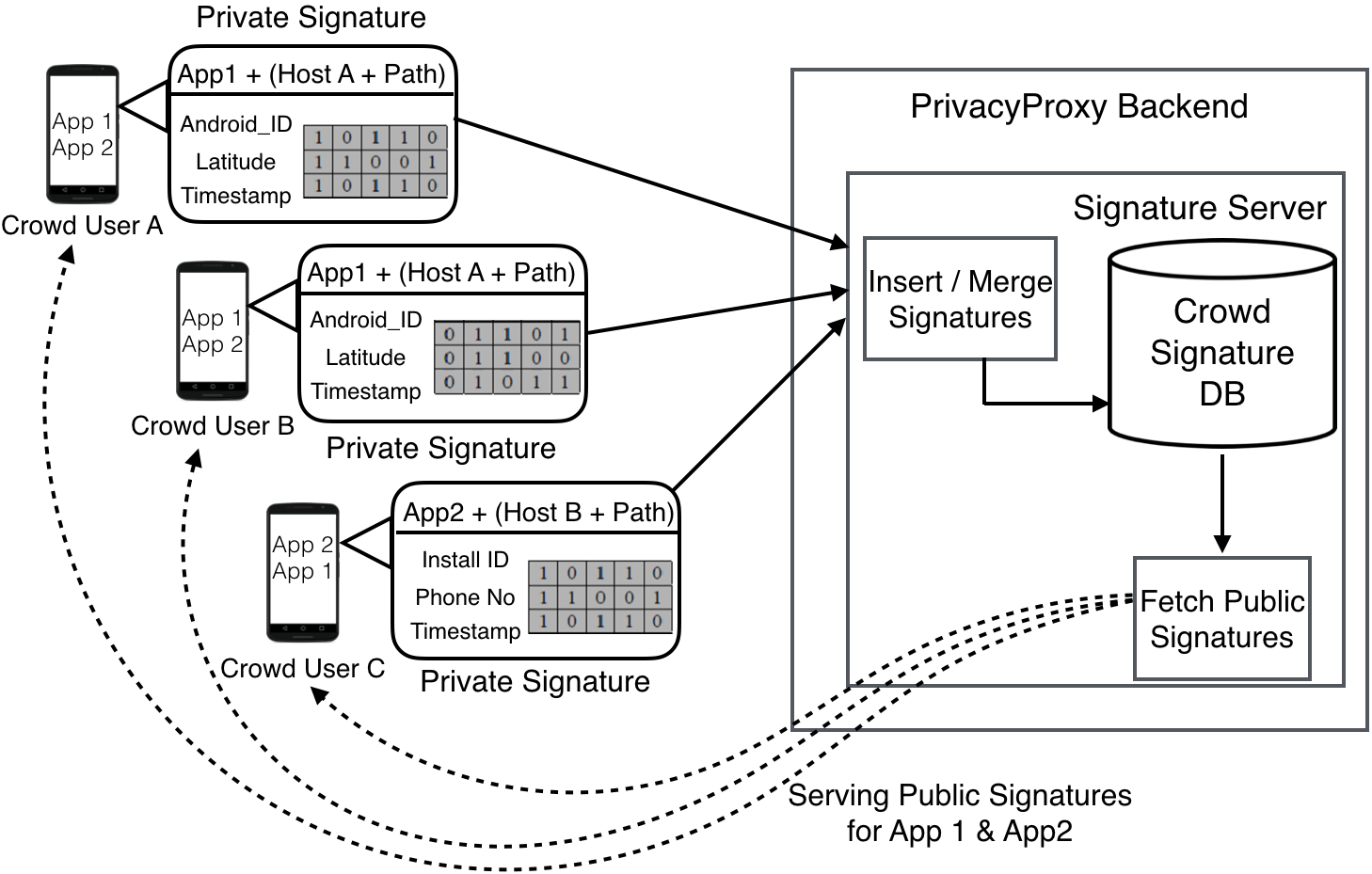}
\caption{
\label{fig:OverallArch}
\textmd{\PP detects likely PII using network analysis and crowdsourcing. Users upload anonymized private signatures of key-value pairs extracted from their network traffic. Signatures with the same SID are merged into a public signature. \PP compares these public signatures with the private signatures to detect PII.}}
\vspace{-3mm}

\end{figure}

\vspace{-1mm}
\subsection{Detecting PII}
\vspace{-3mm}




\begin{figure*}
\centering
\begin{adjustbox}{center, width=(\columnwidth-10pt) * 2}
\begin{tabular}{ccc}
 {
   \includegraphics[height=3.5cm,width=5.25cm]{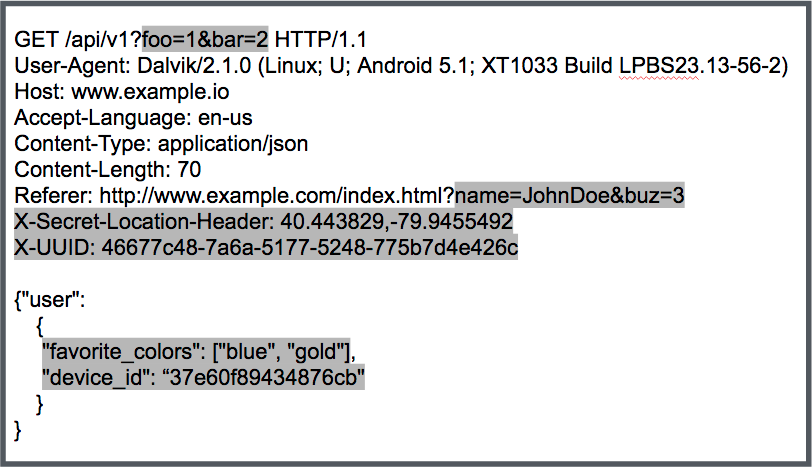}
 } &
 {
   \includegraphics[height=3.5cm,width=5.25cm]{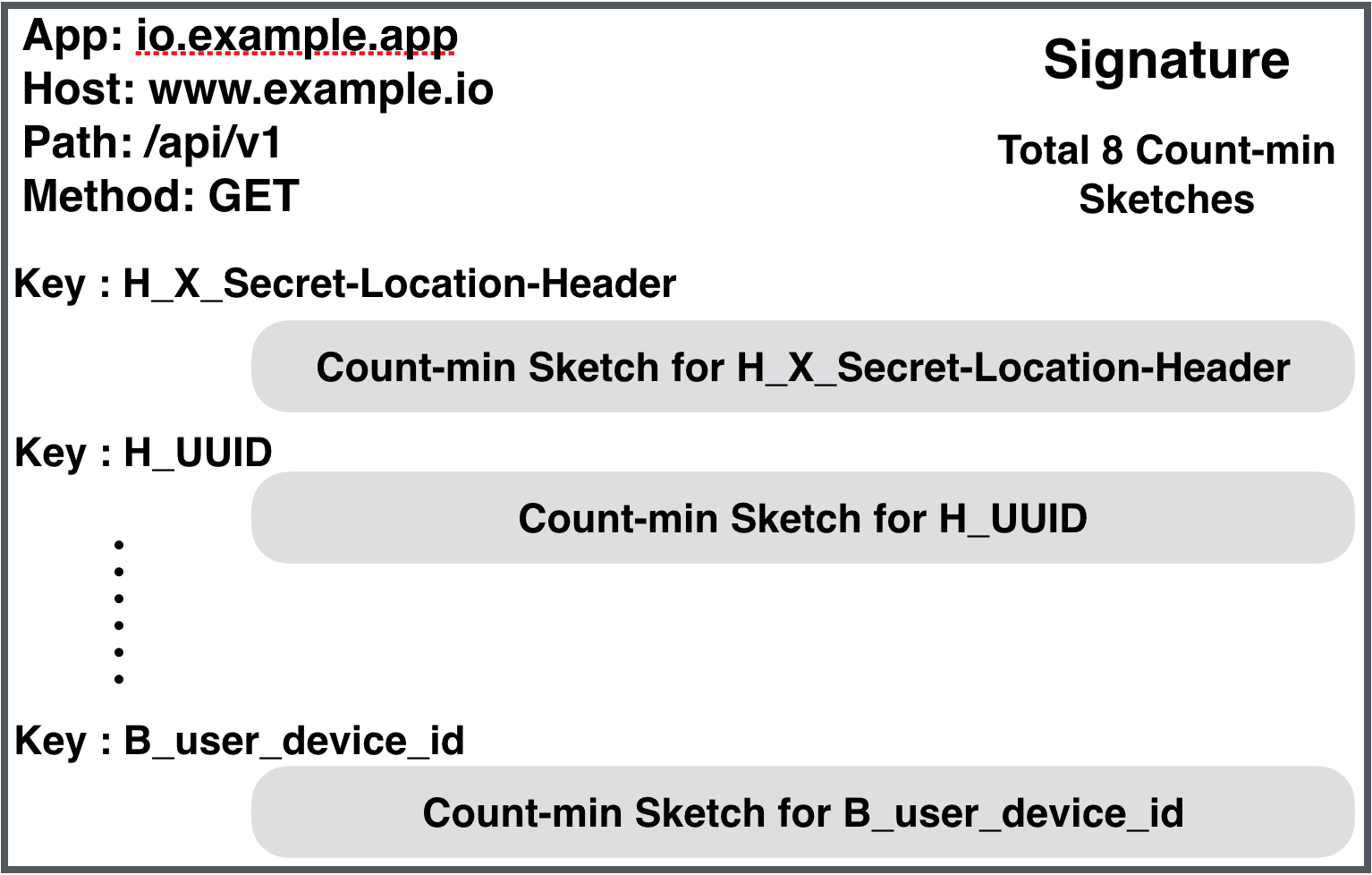}
 } &
 { \includegraphics[height=3.5cm,width=5.25cm]{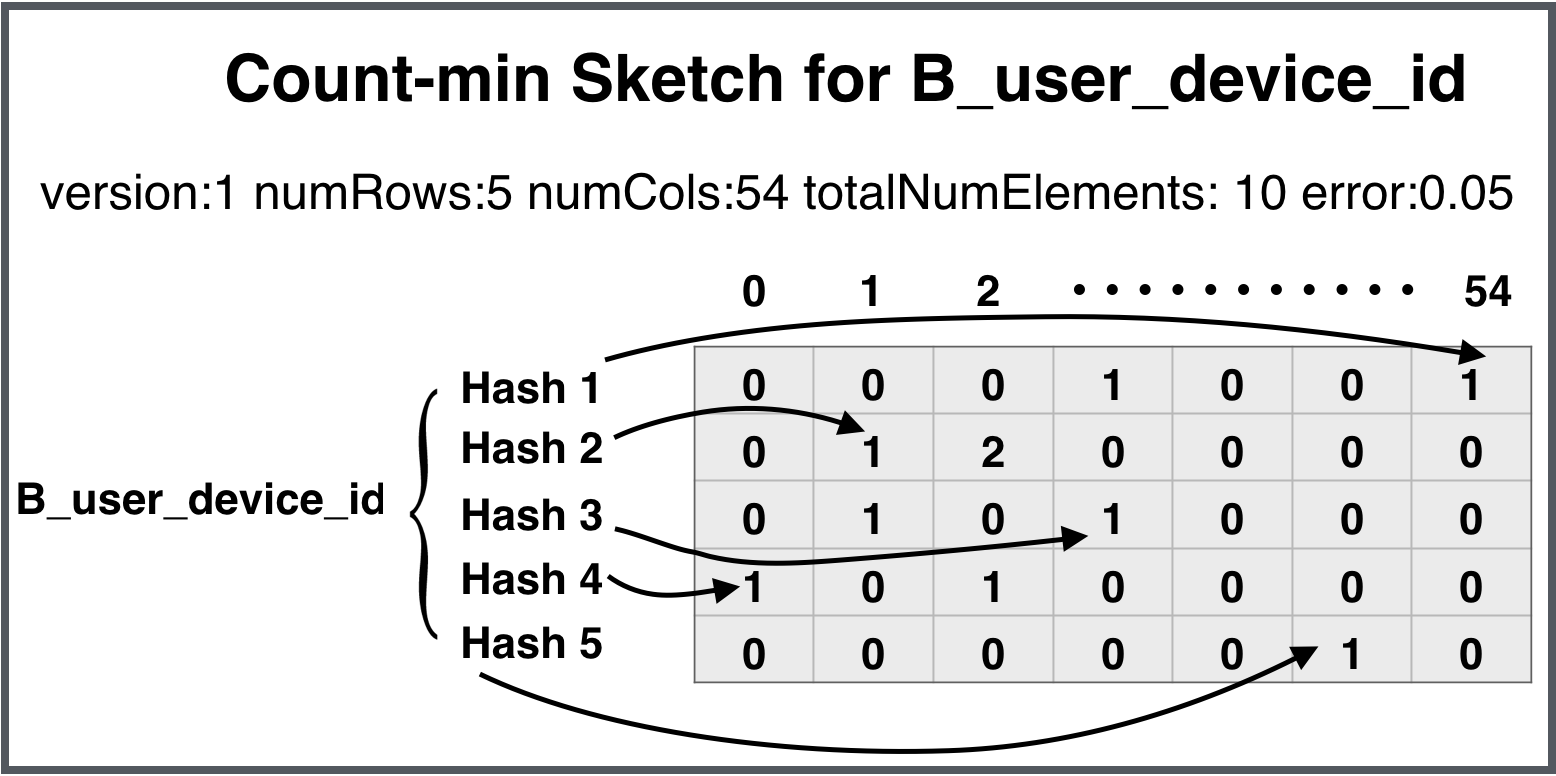}
 } \\
\end{tabular}
\end{adjustbox}
\caption{\textmd{
\label{fig:pii:http}
\label{fig:pii:signature}
\label{fig:pii:sketch}
Signature generation process flow: (a) Our Request Parser looks for key-value pairs in app HTTP requests, looking at request URI query parameters, HTTP headers, and body. Extracted key-value pairs are highlighted. (b) Signature structure for the sample HTTP Request. A signature identifies the app used, destination, URL Path, and method, plus count-min sketch for all keys. c) Count-min sketch signature for key Device\_ID. Count-min sketch is a probabilistic and compact data structure for counting the frequency of different values seen for each key.
}}
\vspace{-3mm}
\end{figure*}

\subsubsection{Extracting Key-Value Pairs} 
\label{sec:arch:extractKV}
After intercepting traffic with the LocalServer, we extract \texttt{Key-Value} pairs. We assume most apps use HTTP(S) RESTful APIs. Prior work found HTTP(S) is a common protocol for client/server communication in Android \cite{A_First_Look_at_Traffic_on_Smartphones}, with over 80\% of apps having structured responses \cite{continella_ndss2017_obfuscation}. However, in our evaluations we observe that close to 100\% of the network requests made by the Top 500 most popular apps use the HTTP(s) protocol.
Figure \ref{fig:pii:http} shows an example request and extracted key-value pairs. We extract all arguments in the request URI. We also extract key-value pairs from HTTP headers, ignoring common headers like ``Accept-Encoding'' and ``Content-Type'', instead looking for uncommon headers like ``X-Secret-Location-Header''. We parse these header values as JSON, XML, or a URL, and if that fails we treat it as text. For the message body we use hints such as the ``Content-type'' header to parse, failing which we try common formats such as JSON, XML, and URL encoded. If that fails, we treat the entire body as a text value and assign a fixed key to it. 
Note, our scheme will not work if apps use a proprietary encoding. 
In our tests, we observed less than 0.6\% of all HTTP requests used a non-standard encoding we could not parse. Our keys have prefixes such as ``U'' for URLs, ``H'' for header, and ``B'' for body to reduce conflicts. In addition, nested keys are flattened to include all the parent keys. 


\subsubsection{Signature Generation from Key-Value Pairs}
\label{sec:arch:signature}
The next step is to detect likely PII based on extracted key-value data. A novel aspect of our technique is to anonymously aggregate contributed data from individual devices, generating crowdsourced signatures that identify which data items (per app) are likely PII. The intuition is that key-values that repeatedly appear in the data streams from only a particular device are likely PII, while those appearing across many devices are not. 

Each device contributes \texttt{signatures} comprised of a \texttt{Signature ID} (SID) and extracted key-value pairs. A Signature ID, or SID is the tuple \texttt{(package-name, app version, method, host, path)}. For instance, in Figure \ref{fig:pii:http}, the HTTP request was sent from version 1.0 of the io.example.app and would have an SID of (io.example.app, 1.0, GET, www.example.io, /api/v1). SIDs let us associate signatures for the same app/requests from different users' devices. The `host' and `path' fields lets us further distinguish the end-point of PII transmission.

The other part of a signature contains all the key-value pairs with the same SID. For each extracted \texttt{key}, we store a count-min sketch \cite{CountMinSketch} which represents the frequencies of all the different \texttt{values} associated with a given key, as shown in Figure \ref{fig:pii:sketch}. \textbf{Count-min Sketch} is like a 2-dimensional Bloom filter that estimates the frequency of an event. Concretely, each count-min sketch is a collection of \(r\) integer arrays of length \(n\) (every element initialized to 0) and \(r\) hash functions. It supports 2 operations: \(increment(v)\) and \(count(v)\). The \(increment\) function takes in a value \(v\) and for each array $\alpha_i ( 1 \leq i \leq r )$, it increments the integer at $\alpha_i[hash_i(v)]$ by one. The count function returns an estimate of the frequency of value v by returning $min_{1\leq i \leq r}(\alpha_i[hash_i(v)])$. 

Count-min sketch is space efficient as it only requires $r \times n$ integers. Given any value, it will either return the correct count or an overestimate, but never an underestimate. It also does not reveal actual values that were inserted, which is good for privacy. Multiple count-min sketches can be trivially combined by doing a element-wise sum over each array. We use SHA-256 as the hash function and we generate \(r\) different hash functions by appending fixed random strings to the value \(v\). Furthermore, for each count-min sketch, we also maintain a counter \(m\) which keeps track of the number of times the increment function was called, regardless of the value used. We chose $r = \lceil \ln \frac{1}{0.01}\rceil = 5 $ and $n = \lceil \frac{e}{0.05}\rceil = 55 $, which theoretically ensures that if the true frequency of \(v\) is $f_v$, then $P[count(v) \leq f_v + 0.05n] \geq 0.99$.

We generate two types of signatures. \textit{Private Signatures} are generated by each user on their device. Each device periodically uploads their private signatures to our \texttt{Signature Server}. Since each private signature only contains the SID and the count-min-sketches but not the actual key-values, it is extremely space efficient and there are minimal privacy concerns. By combining private signatures from different users, the signature server can generate a \textit{Public Signature} which can then be periodically downloaded by devices based on which apps they have. 
To protect user privacy, we don't authenticate users and signatures. However, this opens up the possibility of Sybil attacks, which we discuss later. 


\subsubsection{Identifying Likely PII}
\label{sec:arch:identify}
Again, the intuition here is that a key-value pair that is unique to and repeatedly seen on a single device is likely PII, while a pair common to many devices is not. For example if an app uses a device-based UUID to track users using the key ``UserID'', its value (the UUID) is likely to be the same for requests from the same user, but different across other users of the same app. 


Let $S_{private}$ and $S_{public}$ represent the private and public signature for a particular SID, respectively. 
Let $F (S, k, v)$ denote the estimated frequency for the value $v$ associated with key $k$ as reported by signature $S$. Likewise, let $C(S, k)$ denote the number of values (not necessarily unique) that have been inserted into signature $S$ for key $k$. For each key-value pair $(k, v)$ extracted from the request, let $P_{private}(k, v)$ be the estimated probability of seeing the pair in the user's own requests and $P_{public}(k, v)$ be the estimated probability of seeing the pair across all users' requests. We can then calculate: 
\vspace{-2mm}
$$P_{private}(k, v) = \frac{F(S_{private},k,v)}{C(S_{private},k)}$$ 

\vspace{-2mm}
We can do the same for $P_{public}(k, v)$.
Let us consider the PII analysis for a given key-value pair $(k,v)$. For brevity, we'll use $P_{private}$ to refer to $P_{private}(k, v)$ and $P_{public}$ to refer to $P_{public}(k, v)$. Let $T$ be a tunable threshold between 0 and 1. Using these definitions, we can classify $(k, v)$ into 3 categories:


\setdefaultleftmargin{0pt}{}{}{}{}{}

\begin{enumerate}
  \item \textbf{Application Constant}
\vspace{-2mm}
  $$P_{private} \geq T \quad and \quad P_{private} \leq P_{public}$$ 
  
\vspace{-2mm}
Values in this category are present in the requests of the user as well as others. As a result, they are likely to be the same for all the requests with the same SID. A good example would be developer IDs used by ad libraries to identify the apps that made the ad request.


 \item \textbf{Context-Sensitive Data} 
\vspace{-2mm}
 $$P_{private} < T$$ 
 
\vspace{-2mm}
These values are different for different requests from a user and are unlikely to be PII or app constants. There are two likely explanations. The first are values dynamically generated by the app, e.g. timestamps, random numbers used by libraries such as Google Analytics to prevent caching \cite{GoogleAnalyticsTrackingCode}, and app-specific checksums. The second are values associated with the user's actions. Examples include number of minutes the screen is on or the user's GPS location if the user frequently moves.


\item \textbf{Likely Personally Identifiable Information}
\label{sec:PII_Experiment_def}
\vspace{-2mm}
  $$P_{private} \geq T \quad and \quad P_{private} > P_{public}$$

\vspace{-2mm}
To be classified as PII, a value needs to be common for a single user and rare for others. The selection of parameter $T$ is flexible, although it should be set such that it prevents context-sensitive values from being misclassified as PII or app constants.
Due to the properties of count-min sketch, we will never underestimate $P_{private}$ and $P_{public}$. Along with theoretical guarantees on the bounds of $count(v)$, our PII detection algorithm is quite conservative in determining whether a given value is PII or not. 
We err somewhat on the side of higher false positives, under the assumption that more network data will reduce likelihood of false positives. However, if we err on the true negative side, we risk PII being leaked. 
\end{enumerate}


\PP relies on crowd contributions of private signatures. Even though we only need two users using the same app to start making inferences about likely PII, the true strength of our system is realized as the number of users increases, which improves both signature coverage and accuracy.


If only one network request for a given SID is observed, $P_{private}$ will be 1. As a result, the keys might be incorrectly marked as likely PII until more instances are observed. The higher the number of network requests \PP observes for the same SID, the higher is the likelihood of correctly classifying them as PIIs. To account for this behavior, we introduced a Confidence Threshold $CT$, which dictates the minimum number of times a SID must be seen before \PP starts classifying it. This confidence threshold can also be configured by the users depending upon their tolerance for false positives and false negatives. We evaluate the effect of varying confidence thresholds in Section \ref{subsec:eval:extensivetraining}.

\vspace{-1mm}
\subsection{Mitigating PII Leaks}
\label{sec:arch:mitigatingPII}

\textbf{Filtering PII:} Our base system does not require any user involvement in labeling PIIs, and as we show in our evaluation can still identify PIIs reasonably accurately. In addition, we have an optional functionality letting users explicitly confirm whether a value marked as PII is correct or not. We are conservative, and only after user confirmation do we filter the PII by identifying the PII's structure using regular expressions, and replacing it with anonymous values. For example, we can detect MAC addresses, hashes, IPv4 and IPv6 addresses, phone number, Social Security Number, and email addresses and use static replacements for each of them. Currently these fake values are the same for all users. If the value to be filtered does not match any regex, we replace it with a set of zeros maintaining the exact length of the original value. 

\textbf{Impact of Filtering on Apps:} 
Filtering requests to replace PII with non-PII is one feature of \PP. However, this may impact app functionality or even cause the app to crash. 
If filtering a PII from an app breaks it (e.g. response code is not a HTTP/200 OK), \PP sends the resulting response code (e.g. HTTP/400 or HTTP/500) to the server. If response codes for multiple users (for the same key-value pair) indicates that filtering broke the app, we add it to a global app-crash whitelist and display a warning to the user about filtering that particular PII. In addition, the user can provide feedback if filtering a particular key-value pair made an app unusable. This information is also sent to the server, and if multiple users report the same key-value pair for a particular app, it can be added to a global app-unusable whitelist. This is a conservative approach since it may leak PII, but keeps a good user experience.

\section{Implementation}
\label{sec:impl}
Here, we describe our implementation of \PP on Android.
Note that our approach can be ported to any smartphone OS that exposes a VPN API. We plan to make the source code of \PP available once the results have been published.

\vspace{-2mm}
\subsection{\PP Android App}

We use PrivacyGuard's FakeVPN and LocalServer to intercept traffic and perform MITM attack on SSL/TLS traffic. Both PrivacyGuard \cite{PrivacyGuard} and \PP use Android's \texttt{VPNService} API (added in API Level 14), which is intended for 3rd party VPN apps. These VPN-based apps require the BIND\_VPN\_SERVICE permission. The Android OS explicitly shows a warning to users if an app can monitor their network traffic.

\PP works on devices running API 15 (Android 4.3) to API 23 (Android 6.0). We ported PrivacyGuard to Android Studio 2.2.2 and modified 3324 lines of code for new functionality in \PP, e.g. parsing HTTP requests. The rest of our app (processing logic, UI, XML, and other design elements) added another 18,475 lines of code.
\begin{figure}
\centering
\includegraphics[width=\columnwidth-10pt]{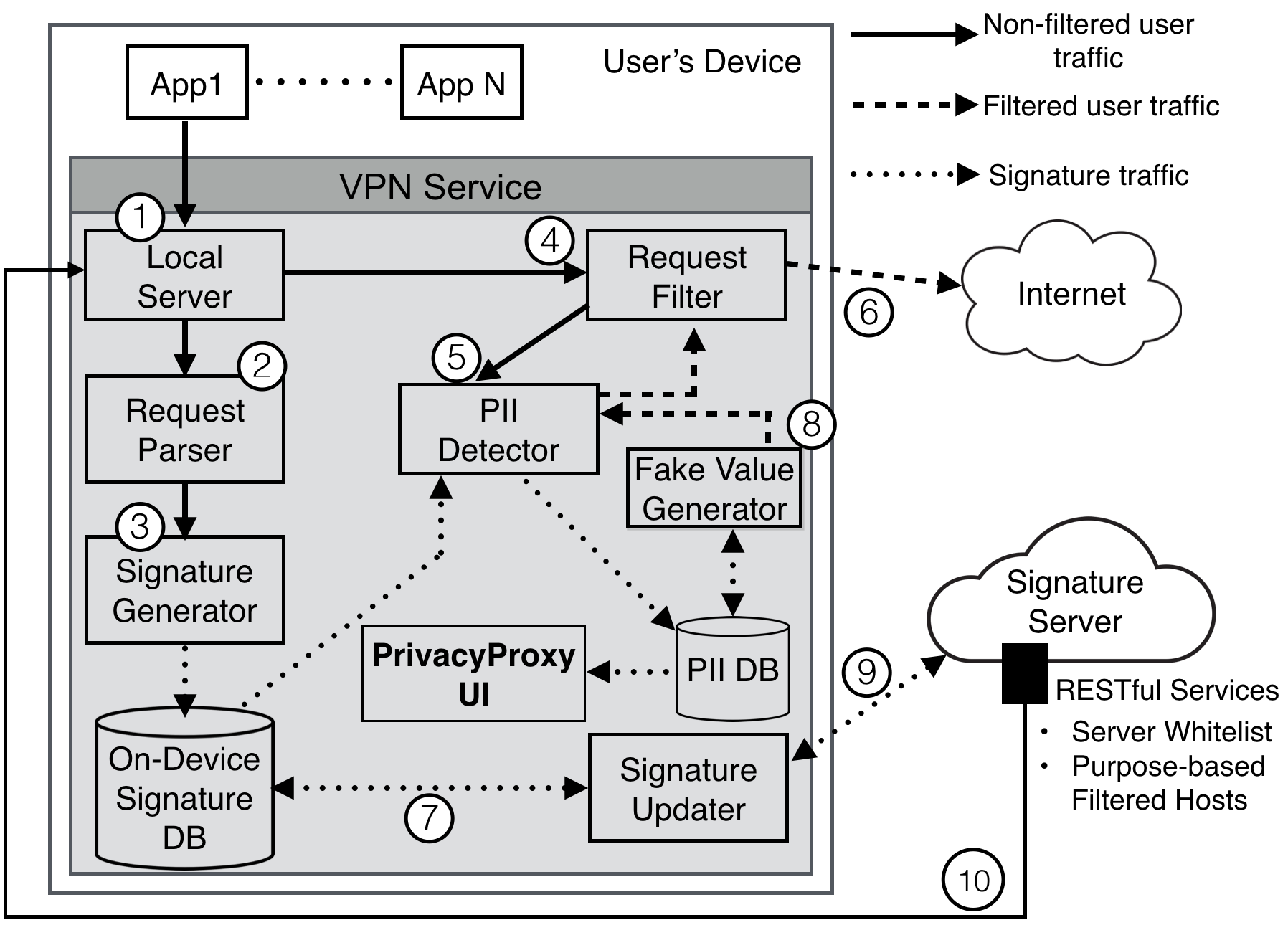}
\caption{
\label{fig:PPAppimpl}
\label{PPServer}
\textmd{Different components and processing steps of \PP, including (1) packet capture, (2) request parsing, (3) signature generation, (4) Request filtering/marking, (5) PII detection, (6) forwarding to destination, (7) upload/download signatures, (8) generating fake data for values marked for filtering, (9) getting public signatures, and (10) host-purpose mapping, server-whitelist from the Signature Server.}}
\vspace{-4mm}
\end{figure}

\begin{figure*}
\centering
\begin{tabular}{cccc}
\tcbox[sharp corners,top=-3pt,left=-3pt,right=-3pt,bottom=-3pt]{{\includegraphics[width = 1.4in]{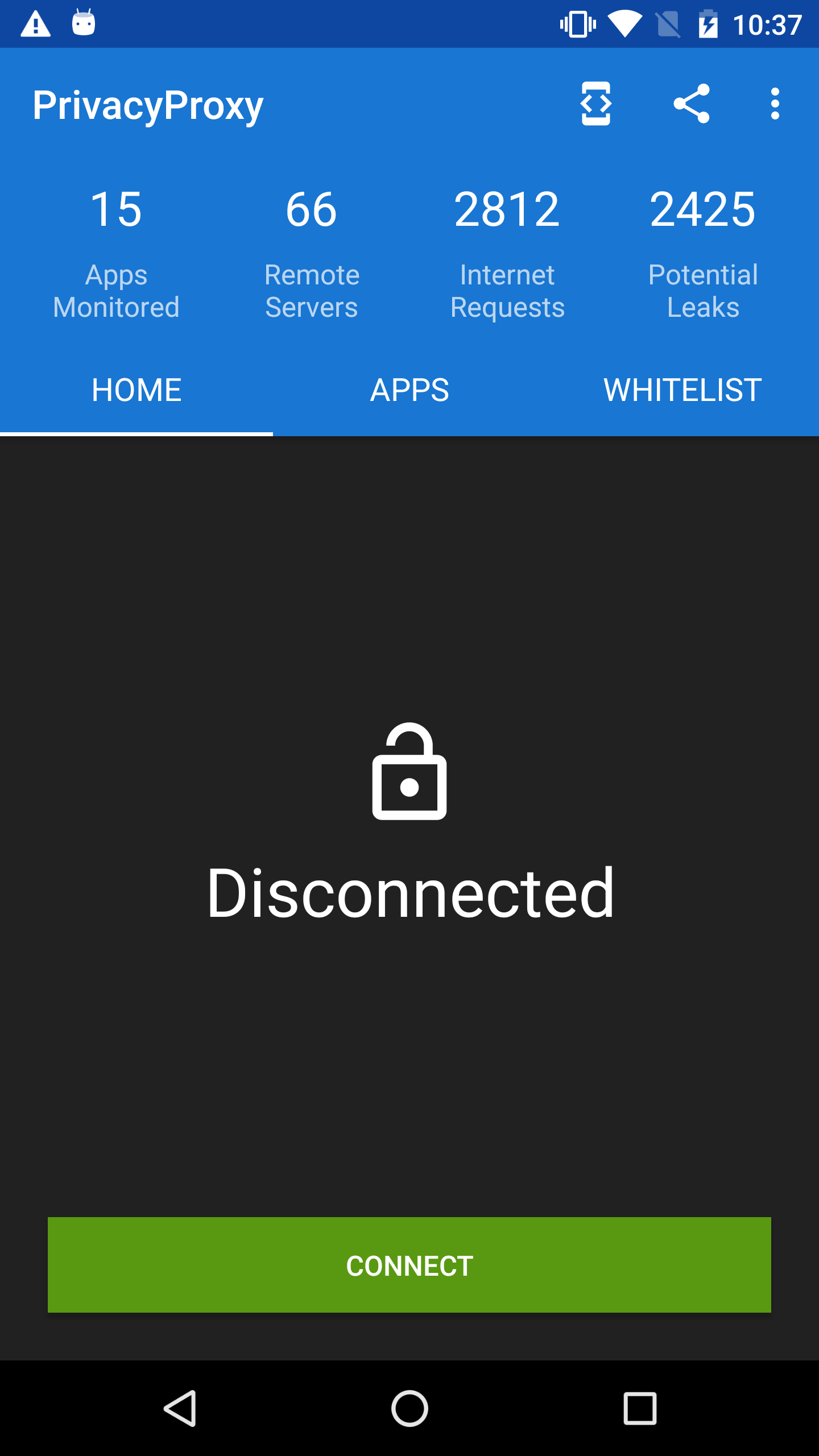}}} &
\tcbox[sharp corners,top=-3pt,left=-3pt,right=-3pt,bottom=-3pt]{{\includegraphics[width = 1.4in]{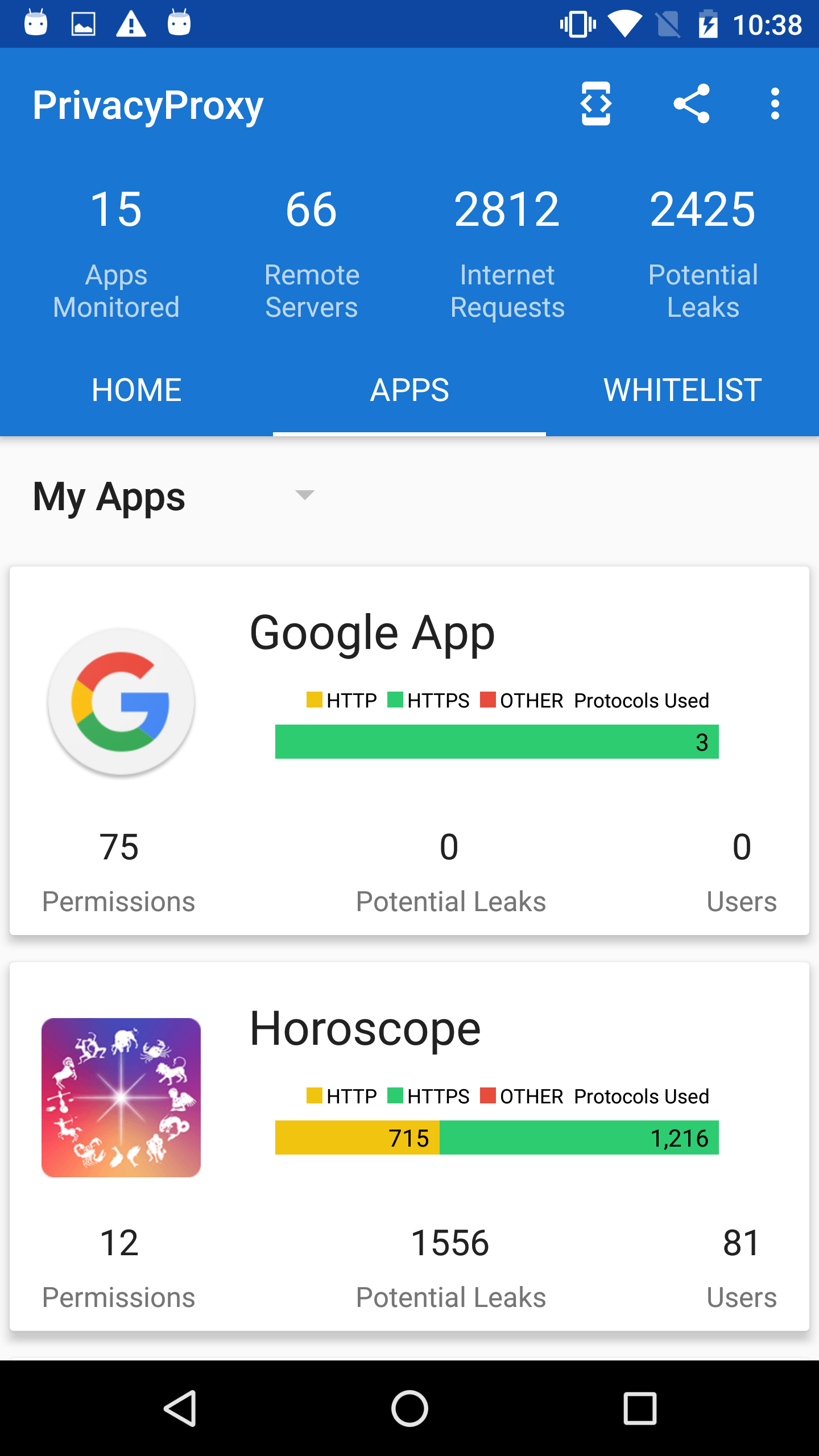}}} &
\tcbox[sharp corners,top=-3pt,left=-3pt,right=-3pt,bottom=-3pt]{{\includegraphics[width = 1.4in]{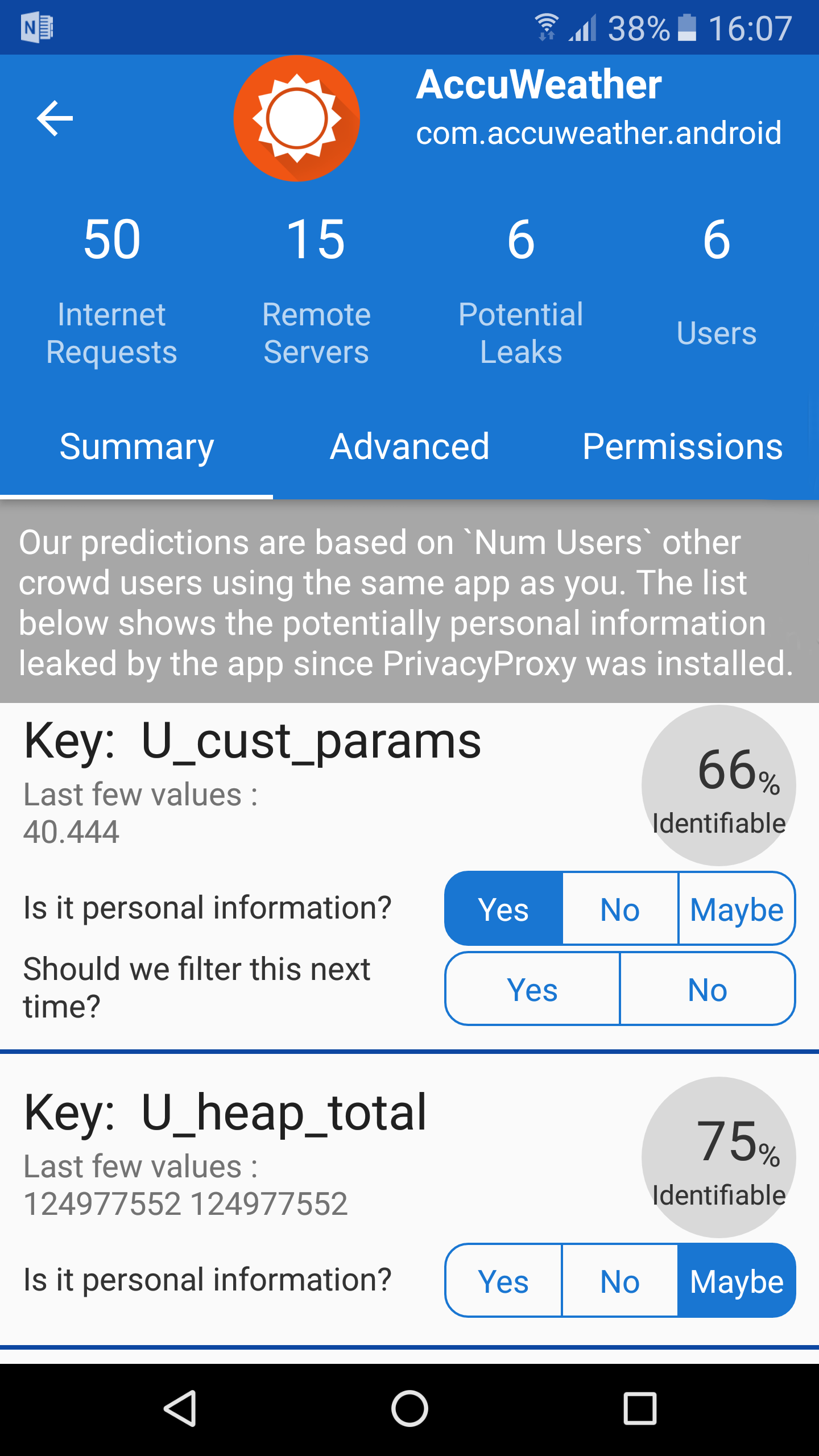}}} &
\tcbox[sharp corners,top=-3pt,left=-3pt,right=-3pt,bottom=-3pt]{{\includegraphics[width = 1.4in]{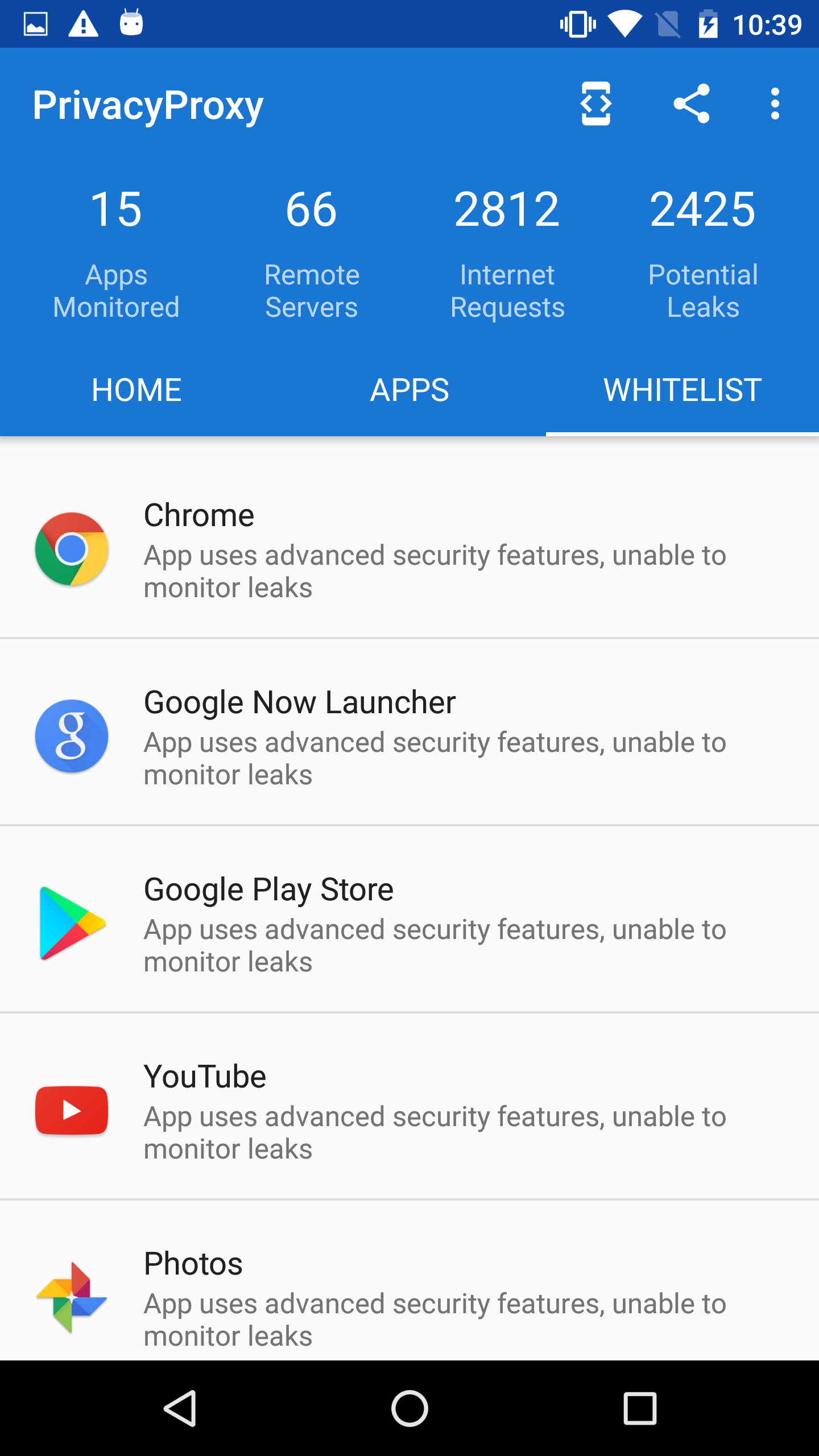}}} \\
\end{tabular}
\vspace{-5mm}
\caption{
\label{fig:PPAppScreenShot}
\textmd{Screenshots of the PrivacyProxy app from left to right: (a) Home screen showing summary information, (b) a list of monitored apps along with summary information for each app, (c) drill-down information for a specific app, showing key-value pairs that have been sent off the device, (d) the whitelist screen}}
\vspace{-2mm}
\end{figure*}

Figure \ref{PPServer} shows the overall flow of data. App requests are intercepted by the FakeVPN Service and sent to the LocalServer (Step 1). Next, our \texttt{Request Parser} (Step 2) attempts to parse each request using several known encodings (currently JSON, XML, and URL) and if not marks it as plain text. The output of this step is a set of key-value pairs. Next, our \texttt{Signature Generator} (Step 3) generates private signatures for the \texttt{(package-name, app version, method, host, path)} tuples using count-min sketch, as explained in Section \ref{sec:sysarch}. These signatures are stored in a \texttt{Signature DB} on the device. The LocalServer in parallel also sends the request to the \texttt{Request Filter} (Step 4), which uses the \texttt{PII Detector} (Step 5) to look for PII by comparing private signatures on the device with the public signatures received from the \PP server. Any detected PII (key, values) are also stored locally on the device in a PII DB, along with generated fake values that they can be replaced with (Step 8). 
Detected PII are shown to the user in the \PP UI, where they can label which key-values are actually PII and can choose to filter out these values. Our app stores these decisions in the PII DB and also logs it in an on-device database. 
Based on the user's decision, either the original app request is unmodified, or PII in the request are modified by replacing original values with anonymized ones, and sent over the network (Step 6).

The private signatures stored on the on-device Signature DB are periodically uploaded by a separate \texttt{Signature Updater} to the Signature Server (Step 7). The signature updater uploads signatures opportunistically, preferring uploading over WiFi and when the phone is plugged in. The signature server processes these private signatures and sends back public signatures. The signature updater stores a local copy of the public signatures for future use. This is done in a battery conserving manner since the \PP signature database only requires weak consistency (Step 9). In addition, the \PP Signature server has a RESTful interface to fetch the server side whitelists (Step 10).



Figure \ref{fig:PPAppScreenShot} shows the UI for the \PP app. The home screen (left) presents an overview of \PP. Users can toggle the VPN functionality and see the total \#apps monitored, \#pieces of information sent to remote servers, \#instances of PII detected, and \#hosts that apps on their device have contacted. In the My Apps view (center-left), users can see summary info of all non-system apps that have used the network.
Clicking on a particular app shows the detailed app view (center-right), which shows the associated key-value pairs that have been detected and the last few values for each key. Here, users can confirm whether the identified key-value pairs are PII or not or whether they are unsure.
If the user marks as PII, they can also choose to filter that key-value, replacing it with an anonymized value the next time we observe it (shown for the second key in the same screenshot). Finally (right), users can configure settings for \PP. 


\textbf{Performance and Energy Optimization: }
We implemented several optimizations to minimize battery use. First, we turn off \PP's VPN when entering Android's \texttt{DozeMode}. While in DozeMode, background services sync and transfer data only during specific time windows. 
Based on our empirical measurements, the VPN service consumes resources even if no network traffic is monitored, and this optimization reduces this overhead. 

Second, we use Android's VPN bypass functionality \cite{WhitelistApps} 
to whitelist certain apps so their traffic does not pass through the VPN. We use this feature to skip apps e.g. email or ones that use SSL certificate pinning to prevent MITM (e.g. Facebook and WhatsApp ). 


Third, we have a \textbf{FastPath} option which helps disable VPN interception and packet processing on a per-app basis based on prior app behavior. Apps are added to the FastPath locally on the device and have no connection with our global whitelist. If we do not detect any PII leak after a certain number of requests (randomly chosen to be between 500 and 2000, but configurable) from a particular app, we add that app to the FastPath and do not inspect requests from that app/version. Consequently, apps on the FastPath are dependent on each user's request and may vary from user to user. However, if an app tries to contact a \textit{new host and path} combination we remove it from the FastPath and start inspecting all its requests again. Moreover, whenever \PP is started, we randomly remove a subset of apps from FastPath and enable sampling their requests again (currently set to 200 requests) to make sure the app is continuing to not send out any PII, and to help mitigate Sybil attacks.

\vspace{-2mm}
\subsection{\PP Signature Server}
\vspace{-1mm}

The \texttt{Signature Server} (SS)
combines private signature uploads from clients and updates public signatures for associated signature IDs. These public signatures are stored in the SS database and is periodically downloaded by \PP clients. Count-min sketch makes combining signatures trivial. In addition, each count-min sketch in a signature can be updated independently, so we don't have to worry about users having slightly different signatures with the same SID. Our SS is implemented in GO, supports HTTPS and authentication, and is around 1,350 lines of GO code. 
To minimize storage and bandwidth, we use Google's Protocol Buffer library to serialize all data transfer. As mentioned earlier, our count-min sketch has $r$ rows (currently 5) and $n$ columns (currently 55) see Figure \ref{fig:pii:sketch} (c). We also use 64 bit integers. This means that each key only takes up $r \times n \times 8$ = 2200 bytes regardless of the number of values. 
To protect user privacy, we do not store the IP addresses or any other identifiers of \PP clients sending us private signatures. 


\textbf{Using the crowd to mitigate user burden:}
Similar to the FastPath optimization earlier, we have also implemented a server-side \texttt{WhiteListing} feature that can provide hints to \PP clients to reduce performance overhead. The intuition is that if there are enough users for a particular app/version (currently set to 5 users) and if we have not detected PII in any of their requests, it's unlikely that the app leaks any PII. For this feature, we aggregate data on the number of requests seen and the number of potential PII flagged by \PP for a particular app/version. Note, we do not need the actual PII to be shared with our server but only the \textit{number} of PII detected and flagged as TP/FP for each app/version. 
If we conclude that a particular app does not leak PII, we add it to our whitelist. \PP client apps can download this whitelist and add them to the VPN bypass method (Step 10 Figure \ref{fig:PPAppimpl}). The advantage for new users is that they will not observe any performance overhead of \PP processing data from these apps. It is possible for an app developer to game the system by not accessing any PII at first. This case can be handled by having a random and small subset of app requests being sampled even if they are in the whitelist, though we have not implemented this feature yet. 

\vspace{-1mm}

\section{Evaluation}
\label{sec:Evaluation}

\begin{table*}
\caption{Detection efficiency in the controlled experiment and field deployment. Rows 2 and 3 show the performance of \PP for the \totalAppsForMoreTraining apps which had maximum number of false positives after 1 and 7 rounds of training runs.} 
\label{tab:precision}
\centering
\begin{tabular}{|l|l|l|l|l|l|l|l|l|}
\hline
\textbf{Row} & \textbf{Scenario}                     & \textbf{Apps} &\textbf{TP}		& \textbf{FP} 		& \textbf{FN} 	& \textbf{Precision} 	& \textbf{Recall} & \textbf{F1 Score} \\ \hline
\hline
1 & Controlled Experiment                 & \totalAppsRunOnDroidbot & \TPControlled	& \FPControlled		& \FNControlled	& \PrecisionControlled	& \RecallControlled & \FScoreControlled		\\ \hline
\hline
2 & Extensive Training - Round 1 & \totalAppsForMoreTraining & \TPAdditionalTrainingStart & \FPAdditionalTrainingStart & \FNAdditionalTrainingStart & \PrecisionAdditionalTrainingStart & \RecallAdditionalTrainingStart & \FScoreAdditionalTrainingStart\\ \hline
3 & Extensive Training - Round 7 &\totalAppsForMoreTraining & \TPAdditionalTraining & \FPAdditionalTraining & \FNAdditionalTraining & \PrecisionAdditionalTraining & \RecallAdditionalTraining & \FScoreAdditionalTraining 	\\ \hline
\hline
4 & Field Study                      & 75 & \TPFieldStudy & \FPFieldStudy & \FNFieldStudy & \PrecisionFieldStudy & \RecallFieldStudy & \FScoreFieldStudy	\\ \hline
\end{tabular}
\end{table*}



We evaluated \PP in five different contexts: an initial controlled experiment, an extensive training experiment, a small field study, a user survey from the field study, and a comparison of \PP with two other related systems - Recon \cite{ReCon} and Haystack \cite{HayStack}. 

These evaluations highlight several key findings. First, we show \PP accurately detects PII in our initial controlled experiment (Section \ref{sec:ControlledExperiment}) with 500 most popular apps (Precision \PrecisionControlled, Recall \RecallControlled, F1-score \FScoreControlled). Second, based on our intuition that additional network requests from the same app will increase accuracy, we show  PII detection accuracy indeed improves (Precision \PrecisionAdditionalTraining, Recall \RecallAdditionalTraining, F1-score \FScoreAdditionalTraining) for the 40 worst performing apps with the most false positives (Section \ref{subsec:eval:extensivetraining}). Third we show in our field study, with 18 participants and users who dicovered \PP organically from the Play Store, that we achieve a precision of \PrecisionFieldStudy and a recall of \RecallFieldStudy (Section \ref{sec:fieldStudy}). Fourth, based on our user survey, we show that \UsersWithNoBatteryLoss of the users in the survey found no perceptible change in battery life and \UsersWithNoNwPerformanceLoss users found no change in network performance (Section \ref{subsec:eval:usersurvey}). Finally, we compare \PP with two prior works, and show that for a set of 15 apps \PP finds significantly more types of PII (89 PIIs) than both Recon (57 PIIs) and HayStack (36 PIIs) respectively (Section \ref{subsec:eval:comparerecon}).


\vspace{2mm}
\subsection{Initial Controlled Experiment}
\label{sec:ControlledExperiment}
\vspace{-1mm}
\textbf{Experiment Design}: 
We selected the top \totalApps free apps from Google Play retrieved on March 13th, 2018. From this set, we were able to evaluate \totalAppsRunOnDroidbot apps. Among the remaining 48 apps,  \appswithCertificatePinning apps used certificate-pinning on all  network requests, \appswithLogin required information such as bank accounts and phone numbers for account creation, and \appswithnonetwork made no network requests at all. We used Nexus 5X running Android 6.0 for these experiments.

Each \textit{training run} for an app consisted of installing the app, exploring it for 150 seconds, and then uninstalling it. We used DroidBot \cite{DroidBot} to perform 4 of these training runs (each 150 seconds) for each app on 4 phones with \PP. Next, we used the same automated script on a fifth phone as a \textit{test run} to evaluate the detected PIIs. The test run was conducted after the training phones had uploaded their signatures to the server.
Note, while exploring each app for longer and doing even more runs would improve coverage, we empirically determined that 150 seconds and 4 runs were good enough to capture a good set of network requests. 

While Droidbot was configured to interact with apps the same way across runs, there were inevitable variations due to non-deterministic behavior in apps,  changes in the environment, or timing issues. For instance, when Droidbot clicks a button, the desired function may or may not complete successfully depending on the network conditions. Similarly, Droidbot's interactions can differ as some apps change the phone environment by turning on accessibility settings, changing the language of the phone, turning off the WiFi, and so on. 

We manually label each key-value pair as potential PII or not for all requests (Total of \PIILeaksControlled unique requests), and then compare them against \PP's classification to calculate precision and recall. We are conservative in our manual labeling of PIIs in that we marked a value as a non-PII if we were not sure about its category, leading to higher false positives. We used a probability threshold of $T = \privateProbThreshold$ based on empirical data. To prevent inflation of the number of detected PII/non-PII values, we only examine unique (SID, key, value) tuples and not multiple instances of the same tuple.


\textbf{Experiment Results}:
The training runs generated \totalPublicSignaturesControlled public signatures from the \totalAppsRunOnDroidbot apps. There were \PIILeaksControlled unique key-value pairs collected from the test device. Of the key-values that \PP identified as PII, we identified \TPControlled True Positives, \FPControlled False Positives, \TNControlled True Negatives and \FNControlled False Negatives (Table \ref{tab:precision}), yielding a precision of \PrecisionControlled. Due to limitations of the coverage of any automated testing framework, we cannot guarantee that we have triggered all possible PII leaks from the apps. Therefore, we calculated the false negatives considering only the network requests we saw in our experiments. 

After manually analyzing false positives, we categorized them into four categories. First, \PP mistakenly detects some values as PII because in certain apps, the URL paths (which \PP uses as signature ID) had variables, e.g. \texttt{api.geo.kontagent.net/} \newline \texttt{api/v1/11828d55da4547b5b7bd85eef33c38a0/cpu/}. \newline 
Since, these variable strings form a part of signature ID, \PP rarely sees enough instances of the request to classify it properly. Second, \PP is unable to correctly classify resource paths on servers (e.g. for an image) sent by the apps without sufficient number of users. Third, values which do not seem to have a semantic meaning may or may not be PII. We conservatively label all such values as not PII for ground truth. Finally, values belonging to network requests made only one time. These values have a private probability of 1 and as discussed previously, \PP is unable to correctly classify such values without more requests.

Interestingly, the Brightest Flashlight LED app transmitted the SHA hash of the UDID. Apps like Draw.ly, Drink Water Reminder, and Yelp used random hashes as identifiers for the device. In such scenarios, regex-based approaches can easily miss PII if we do not have a priori knowledge. Additionally, users may not have sufficient knowledge to label values like these as PII. Lastly, the current permission-based controls in Android would not be able to control such PII leaks.

\vspace{-2mm}
\subsection{Extensive Training Experiment}
\label{subsec:eval:extensivetraining}
\vspace{-1mm}
	Since many of the False Positives in the controlled experiment could be due to not seeing enough network requests with the same SID, our hypothesis was that the precision of \PP's classification will increase as more instances of these requests are observed. 

\textbf{Experiment Design:}
    We tested this hypothesis on a set of \totalAppsForMoreTraining apps for which \PP had most instances of False Positives in the classification of PII during the controlled experiment \ref{sec:ControlledExperiment}. The experimental setup was similar to the controlled experiment, with the duration of each training and testing run increased to 750 seconds. A test run was conducted after each training run was completed and the private signatures had been uploaded to the server. Apps were only installed once at the beginning of the experiment and persisted throughout the experiment. For this experiment, we started from a clean database without using any data from the prior controlled experiment (Section \ref{sec:ControlledExperiment}). 

We also evaluate how different confidence thresholds affect precision and recall. As described at the end of Section \ref{sec:arch:identify}, the confidence threshold ($CT$) dictates the minimum number of times a (SID, key) pair must be seen before it is classified as PII/non-PII. Intuitively, a lower value will have higher recall since even infrequent values will be classified, while a higher threshold will have better precision since \PP will have higher confidence about a key-value being PII or not.  

\begin{figure}
\centering
\includegraphics[width=1\linewidth]{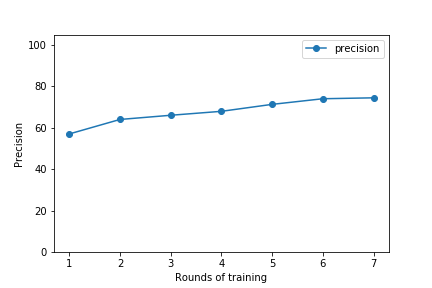}
\caption{Change in precision of PII detection with increasing training runs for 40 apps with most false positives in \ref{sec:ControlledExperiment}.}
\label{fig:PrecisionVsTraining1}
\vspace{-2mm}
\end{figure}

    
\textbf{Experiment Results:}
   As illustrated in Figure \ref{fig:PrecisionVsTraining1}, the average precision for \PP for these 40 apps increases from \PrecisionAdditionalTrainingStart to \PrecisionAdditionalTraining with additional training runs. The reason for this is that the number of False Positives reduce with longer training and more runs. 

\begin{figure*}
\centering
\begin{subfigure}{.5\textwidth}
  \centering
  \includegraphics[width=1\linewidth]{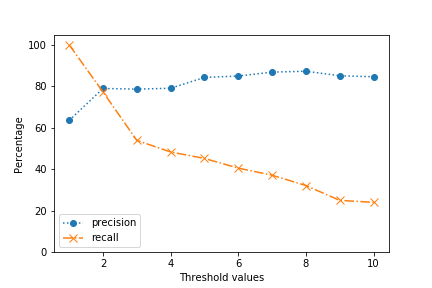}
  \caption{Precision and Recall after 1 round of training}
  \label{fig:TrainingThreshold1}
\end{subfigure}%
\begin{subfigure}{.5\textwidth}
  \centering
  \includegraphics[width=1\linewidth]{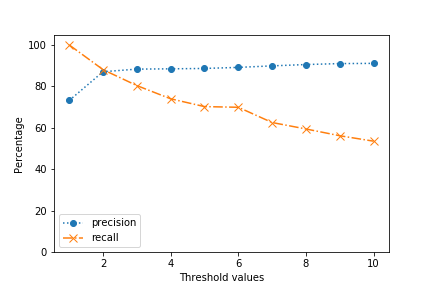}
  \caption{Precision and Recall after 7 rounds of training}
  \label{fig:TrainingThreshold2}
\end{subfigure}
\caption{Effect of confidence threshold on the overall precision and recall of \PP after 1 round and 8 rounds of training. \textmd{Precision and Recall are calculated for 40 apps with the most False Positives in the initial evaluation.}
}
\label{fig:TrainingThreshold}
\end{figure*}

	Figures \ref{fig:TrainingThreshold1} and Figure \ref{fig:TrainingThreshold2} show the effect of varying the confidence threshold on precision and recall, for 1 round of training and 7 rounds of training respectively. The recall drops more dramatically as we increase the threshold values, and the drop off is more dramatic for a single training round as compared to seven rounds. Based on this data, we determine that the near optimal confidence threshold value is 2. However, we note that users can still tune this threshold based on their tolerance for false positives and false negatives.

\vspace{-2mm}
\subsection{Field Study}
\label{sec:fieldStudy}
\vspace{-1mm}    
    
\textbf{Experiment Design:}    
	 We published \PP on the Google Play in Nov 2016. While our field study includes users that found our app organically on the Play store, most of the data collected comes from the 18 participants that we recruited specifically using an online portal (between Nov 2016 and January 2017). Recruited users were asked to install and use \PP for at least 4 days on their own smartphone. This was an IRB-approved study and participants were given a \$20 Amazon Gift Card. During the study, the public signatures were collected in a database on our server, receiving \totalPublicSignaturesFieldStudy public signatures from \totalAppsFieldStudy apps. Next we selected the top 100 apps among these apps, which had the most number of signatures and installed them on a test device. We used this test device to evaluate \PP's PII detection accuracy, obviating the need for actual users to share their original key-value pairs. We interacted with these 100 apps manually on the test device for one minute each. 
     The number of users for this subset of 100 apps ranged from 1 to 10. 54 apps had only 1 user, but 3 apps had 6 users (Moneycontrol, Airdroid, News Break), 1 app had 8 users (Android Backup Service), and 1 app had 10 users (Google Services Framework). 


\textbf{Experiment Results:} 
	\PP captured requests for 75 apps, that communicated with 30 remote hosts. We detect 116 potential PII leaks in 24 apps with \TPFieldStudy being True Positives and \FPFieldStudy False Positives. The precision of classification was \PrecisionFieldStudy (see Row 4, Table \ref{tab:precision}). 


    
\vspace{-2mm}
\subsection{User Survey on Usability}
\label{subsec:eval:usersurvey}
\vspace{-1mm}
In Section \ref{sec:impl}, we described the optimizations we did for reducing energy consumption of \PP. Without these optimizations, the \PP overhead was 30\% higher. To quantify the impact on battery life, we installed the top 20 free apps on our test devices. For the first test, these apps were launched in the background and then the screen of the test device was turned off. In the second test, we used the MonkeyRunner script to open each of these 20 apps one by one and interact with them by performing 100 clicks on the screen. 

	To measure overhead, we calculated the time it took for the battery to go from from 100\% to 85\% with and without \PP. To mimic real world usage, we enabled WiFi, LocationServices, etc on the test device and ran each test thrice to report data. Table \ref{tab:table_batterylife} presents our results. Our data shows that for standby, \PP reduces battery lifetime by 8.6\% - 11.0\%, while in active use the overhead can be up to 14.2\%. Note that prior work has shown that VPNs themselves consume about 10\% battery \cite{rao2013meddle}, so the additional battery consumption due to \PP appears to be small.
    

\textbf{Experiment Design:}
    The participants from the field study (Section \ref{sec:fieldStudy}) were shown an optional in-app survey after 4 days of using the \PP app, to assess their perception of usability and performance. All questions were on a 5-point Likert scale.
    
\textbf{Experiment Results:}
There were 27 respondents, including 18 participants we recruited and others who organically found and used the app and chose to respond to the survey. 85\% of participants reported ``Noticed no perceptible change'' for battery usage, and 67\% for the network performance. Although there is a performance overhead of using \PP, 59\% of users said they did not notice any perceptible change in apps. Furthermore, 54\% of participants reported that the performance overhead was worth the privacy improvement while 31\% were neutral. 

\begin{table*}
  	\caption{Battery Overhead Evaluation. We report the time taken (in minutes) in each case for the battery charge to drop from 100\% to 85\%. \textmd{The percentage in the parenthesis depicts the reduction in battery life (overhead of \PP), which varied from 8.6\% to 14.2\% in our 2 test devices.} Note, for the background case the screen was Off, and for the foreground tests the screen was On and the monkey was used to explore the apps.}
\label{tab:table_batterylife}
\small
\centering
\begin{tabular}{|l|l|l|l|l|}
\hline
\multicolumn{1}{|c|}{\multirow{2}{*}{\textbf{Scenario}}}                      & \multicolumn{2}{l|}{\textbf{Samsung S7 Edge}}                                                                            & \multicolumn{2}{l|}{\textbf{Nexus 5X}}                                                                                 \\ \cline{2-5} 
\multicolumn{1}{|c|}{}                                                        & \textbf{Background}                                            & \textbf{Foreground}                                            & \textbf{Background}                                           & \textbf{Foreground}                                           \\ \hline
\begin{tabular}[c]{@{}l@{}}Apps used without PrivacyProxy\end{tabular}                & 463 min                                                     & 52 min                                                     & 650 min                                                    & 70 min                                                    \\ \hline
\begin{tabular}[c]{@{}l@{}}Apps used with PrivacyProxy\end{tabular}                  & \begin{tabular}[c]{@{}l@{}}416 min \\ (10.2\%)\end{tabular} & \begin{tabular}[c]{@{}l@{}}46 min \\ (11.5\%)\end{tabular} & \begin{tabular}[c]{@{}l@{}}594 min \\ (8.6\%)\end{tabular} & \begin{tabular}[c]{@{}l@{}}61 min\\ (12.8\%)\end{tabular} \\ \hline
\begin{tabular}[c]{@{}l@{}}Apps used with PrivacyProxy and Filtering enabled\end{tabular} & \begin{tabular}[c]{@{}l@{}}412 min\\ (11.0\%)\end{tabular}  & \begin{tabular}[c]{@{}l@{}}45 min \\ (13.4\%)\end{tabular} & \begin{tabular}[c]{@{}l@{}}587 min \\ (9.7\%)\end{tabular} & \begin{tabular}[c]{@{}l@{}}60 min\\ (14.2\%)\end{tabular} \\ \hline
\end{tabular}
\vspace{-3mm}
\end{table*}

\vspace{-3mm}
\subsection{Comparison with Recon and HayStack}
\label{subsec:eval:comparerecon}
\vspace{-1mm}
	We compared \PP with Haystack \cite{HayStack} and Recon \cite{ReCon}. We selected 15 random apps from Top 100 free apps from Google Play Store on July 31st, 2017. We tried to download and use Recon but were unable to get it to work. Therefore, we contacted the authors of Recon, and sent them these 15 apps so that the authors may run their system and send us the PII detections for these apps, which they gracefully did. We installed \PP and Lumen Privacy Monitor (HayStack) on a test device, and then used \PP and HayStack one after the other to report flagged PIIs for the same 15 apps. For \PP, we used 5 phones to generate signatures and then used the test phone to evaluate. 
    
    For brevity, we have included a two representative apps (row1, row2) from our test setup in Table \ref{tab:app_wise_comparision} and we include the PIIs detected by those apps. We also include the total PIIs detected for all 15 apps (row3). Our goal was to evaluate whether there are certain PII that approaches based on user supplied labels (ReCon) and deep packet inspection (Haystack) would miss as compared to \PP and vice versa. As noted in Table \ref{tab:app_wise_comparision}, \PP identified many more correct PII than either approach (True Positives). Considering true positives, \PP found 89 PII while Recon detected 57 and HayStack found 36. Being a more generic approach, \PP has more false positives than the other two approaches while detecting more PII than other specialized approaches. 
    

\begin{table*}
\caption{
  Comparison of \PP, Recon, and HayStack in detecting unknown PII for 15 Apps. 
  \textmd{The first two rows present example apps showcasing the ability of \PP to find non-standard PII as compared to other approaches. We categorize false positives based on the strict definition of PII we use in this paper. For example, Gender and Time-Zone alone cannot identify a person, though can if combined with other values, as discussed in Section \ref{sec:limitation}. }
}
\label{tab:app_wise_comparision}

\small
\centering
\begin{adjustbox}{center, width=(\columnwidth-11mm) * 2}

\begin{tabular}{|l|l|l|l|l|l|}
\hline
\textbf{Row} & \textbf{Package Name} & \textbf{Metric} & \textbf{ReCon} & \textbf{HayStack} & \textbf{PrivacyProxy} \\ \hline
\hline
\multirow{2}{*}{1} & \multirow{2}{*}{com.abtnprojects.ambatana} & TP & \begin{tabular}[c]{@{}l@{}}Advertiser ID, Android ID, \\ Device ID, Email\end{tabular} & \begin{tabular}[c]{@{}l@{}}Account Information, \\ Advertiser ID, Android ID, \\ Hardware ID\end{tabular} & \begin{tabular}[c]{@{}l@{}}Advertiser ID, Device ID, \\ First Launch Date, GAID, Hardware ID, \\ Identity ID, Install Date, UID\end{tabular} \\ \cline{3-6} 
 &  & FP & Location & Build Fingerprint & Local IP, Timestamp \\ \hline
\multirow{2}{*}{2} & \multirow{2}{*}{tv.telepathic.hooked} & TP & Advertiser ID, Android ID & \begin{tabular}[c]{@{}l@{}}Account Information, \\ Android ID\end{tabular} & \begin{tabular}[c]{@{}l@{}}Android ID, AppsFlyerID, Device ID, \\ First Launch Date, GAID, Hardware ID,\\ Identity ID, SessionID, UID\end{tabular} \\ \cline{3-6} 
 &  & FP & - & Build Fingerprint & Date, Timestamp \\ \hline
\hline
\multirow{2}{*}{3} & \multirow{2}{*}{PII from all 15 apps} & TP & 57 & 36 & 89 \\ \cline{3-6} 
 &  & FP & 16 & 18 & 21 \\ \hline
\end{tabular}

\end{adjustbox}
\end{table*}
\section{Discussion and Limitations}

\PP is in theory subject to the cold start problem, where a new user of an app has to wait until we have enough signatures from other users of the same app to have public signatures and to detect PIIs. We also showed that our accuracy of PII detection improves with more data. In practice, one would not start with a clean signature database. For example, the data that we collected by the Droidbot UI automation tool \cite{DroidBot} for the most popular 500 apps running on our training and test devices, and the associated public signatures can itself serve to bootstrap new users. This approach can also be applied periodically on a much larger subset of popular apps. 


\label{sec:discussion:security}

One security concern with \PP is Sybil attacks where a malicious entity attempts to poison our signature database. Sybil attacks have been studied in the context of recommender systems, with a number of proposed defenses \cite{dsybil,Cheng_sigcomm2005_SybilproofMech,Frey_Sybil,Noh_icc13_sybilProtection}. 
For \PP, some options include: (a) gathering a great deal of known good data (e.g. our bootstrapping approach above); (b) determining on the server whether an uploaded signature is honest or not before merging it with public signatures; (c) verifying that honest users are fairly long lasting and upload signatures consistently; (d) providing good PII detection without considering all signatures. DSybil \cite{dsybil} has a similar context to \PP where recommended objects can be good or bad, the lifetime of users is relatively long lived to build trust, and the user aims to find some good objects rather than finding all good objects necessarily. We believe that applying ideas from Dsybil can help \PP mitigate against these attacks. Another mechanism is to compare the frequency estimates from uploaded signatures to those existing in the database. If the uploaded signature differs substantially, we can reduce their relative weight when combining signatures or hold off using them until we see similar signatures for the same SID. We defer exploration of defending against Sybil attacks to future work. 

\label{sec:limitation}
\PP also has some limitations. First, it is not designed to defend against malicious developers. For example, \PP can be evaded by certain kinds of obfuscation. However, it is worth noting that all other existing approaches today would not work under active obfuscation, and that \PP is harder to evade than these existing approaches. 

Second, we currently do not parse the request URI path for PII, as it is not a structured key-value pair. However, we observed that some apps send the username and email as part of the path. We have the path as part of our SID, and so it is possible to add values from the path to our analysis, though it is still an open question as to the best way of using this data. 


 
Third, \PP cannot detect leaks from certificate pinned apps. While certificate pinning improves security by mitigating man-in-the-middle attacks, it also makes it harder for any intermediary to inspect network flows for PII. 
In our pilot evaluations, we skipped any app that used certificate pinning for any network request.  However, upon closer inspection, we discovered that less than 5\% of the top 500 free apps use certificate pinning for every request. We bypass these apps and add them to a global whitelist and notify the user. The rest of the apps which use certificate-pinning do so for only a subset of network requests, still allowing us to monitor the requests not certificate pinned. We note that the future of certificate pinning is unclear. Google announced that pinning will be deprecated in future versions of Chrome \cite{deprecationOfCertificatePinning}, due to challenges with site maintenance and accidentally blocking legitimate visitors. Given certificate pinning's low adoption in the most popular apps and its uncertain future, we feel that this is not a major issue for our work. 

Fourth, \PP may not be able to successfully detect PII if we fail to properly parse the request and extract the key-value pairs. In practice, we did not found many instances of this. 

Fifth, \PP works best for pinpointing unique identifiers that are single values, but less so for other PII. For example, some values by themselves are not highly identifiable, but can be when combined with others, e.g. device type, timestamps, location (see \cite{GenderInPII} for more). 
These values usually belong to three types of identity knowledge that neither \PP nor past work supports well: pattern knowledge, social categorization, and symbols of eligibility \cite{DefinitionOfPII}. 
\PP also does not work well for smartphone sensor data, which can also leak private information \cite{Wang_Mobicom15_MotionLeaksThroughSensors,Roy_Mobisys14_SmartphoneWalkDirection,dey2014accelprint}. 




Finally, a malicious signature server could conduct a dictionary attack revealing the actual values from the hashes. One attack is a known value attack, e.g. precomputing the hash for a specific email address and looking for that hash in uploaded signatures. We currently do not have a way to defend against this attack. 
Another attack is to precompute the hashes of all possible values if the range is small, e.g. all possible birth years or ages. Again, our focus is on unique identifiers, making dictionary attacks difficult since the search space is quite large. Nevertheless, there are some possible ways to mitigate this attack. First, we can add some noise to signatures before they are uploaded to a signature server to make precomputation harder, e.g. adapting techniques similar to those in RAPPOR \cite{RAPPOR_CCS2014}. However, this risks increasing false negatives. Second, we might also only retrieve the counts from the signature server instead of the whole signature, which will limit the information attackers can infer from public signatures and make it much harder to do offline dictionary attacks. Third, we can segregate count-min sketches of different keys to prevent correlation attacks between different values (like inferring what addresses someone goes to). We can then use an encrypted or hashed value of SID and a key to find the right count-min sketch without revealing the count-min sketch of corresponding keys from the same signature. 

With \PP, we opted to push our design to one extreme, seeing how far we could go using our crowd-based approach, using minimal information about and processing on each device. In practice, for an actual deployed system, one could combine our approach with other complementary approaches. For example, one could use regular expressions to quickly find common types of PII, especially for apps with few users, and use our crowd-based approach to find uncommon types of PII for widely used apps. Using regex would also make it easier for \PP to more easily filter out certain cases too, e.g. timestamps of the current time.

\section{Conclusions and Future work}
\label{sec:Conclusions}
\PP is a scalable system for detecting and mitigating leaks of PII from devices running stock Android OS. We use a unique signature structure to efficiently summarize all network requests leaving the user's device without revealing actual content. In addition, we propose a novel crowd-based PII detection algorithm that relies on user-generated signatures instead of heuristics. This approach lets us detect both traditional forms of PII as well as app-specific identifiers. By running on the user's device, we can filter out PII before it leaves the device, and we have minimized the network overhead and battery consumption of our app.

Based on our experimental evaluation, we found that our precision for detecting valid PII increases as more users use \PP. In addition we show that \PP can find several types of PII that other approaches cannot. Furthermore, we show that perceived impact on battery life as well on the network performance is low. 

{\bibliographystyle{IEEEtran}
\bibliography{literature}}

\end{document}